\documentclass[prl,aps,twocolumn]{revtex4}
\usepackage{graphicx}
\usepackage{epstopdf}
\usepackage{latexsym}
\usepackage{amsmath}
\begin{document}
\title{Quantum phase transitions in $d$-wave superconductors}
\author{Matthias Vojta, Ying Zhang, and Subir Sachdev}
\address{Department of Physics, Yale University\\
P.O. Box 208120, New Haven, CT 06520-8120, USA}
\date{July 11, 2000}
\begin{abstract}
Motivated by the strong, low temperature damping of nodal
quasiparticles observed in some cuprate
superconductors,
we study quantum phase transitions in $d_{x^2-y^2}$ superconductors
with a spin-singlet, zero momentum, fermion bilinear order
parameter.
We present a complete,
group-theoretic classification of such transitions into 7 distinct cases
(including cases with nematic order) and analyze
fluctuations by the renormalization group. We find
that only 2, the transitions to $d_{x^2-y^2}+is$ and $d_{x^2-y^2} + i d_{xy}$
pairing, possess stable fixed points with universal damping of
nodal quasiparticles; the latter leaves the gapped
quasiparticles along $(1,0)$, $(0,1)$ essentially undamped.
\end{abstract}
\pacs{PACS numbers:}
\maketitle

\begin{figure}[t]
\centering
\includegraphics[width=3.3in]{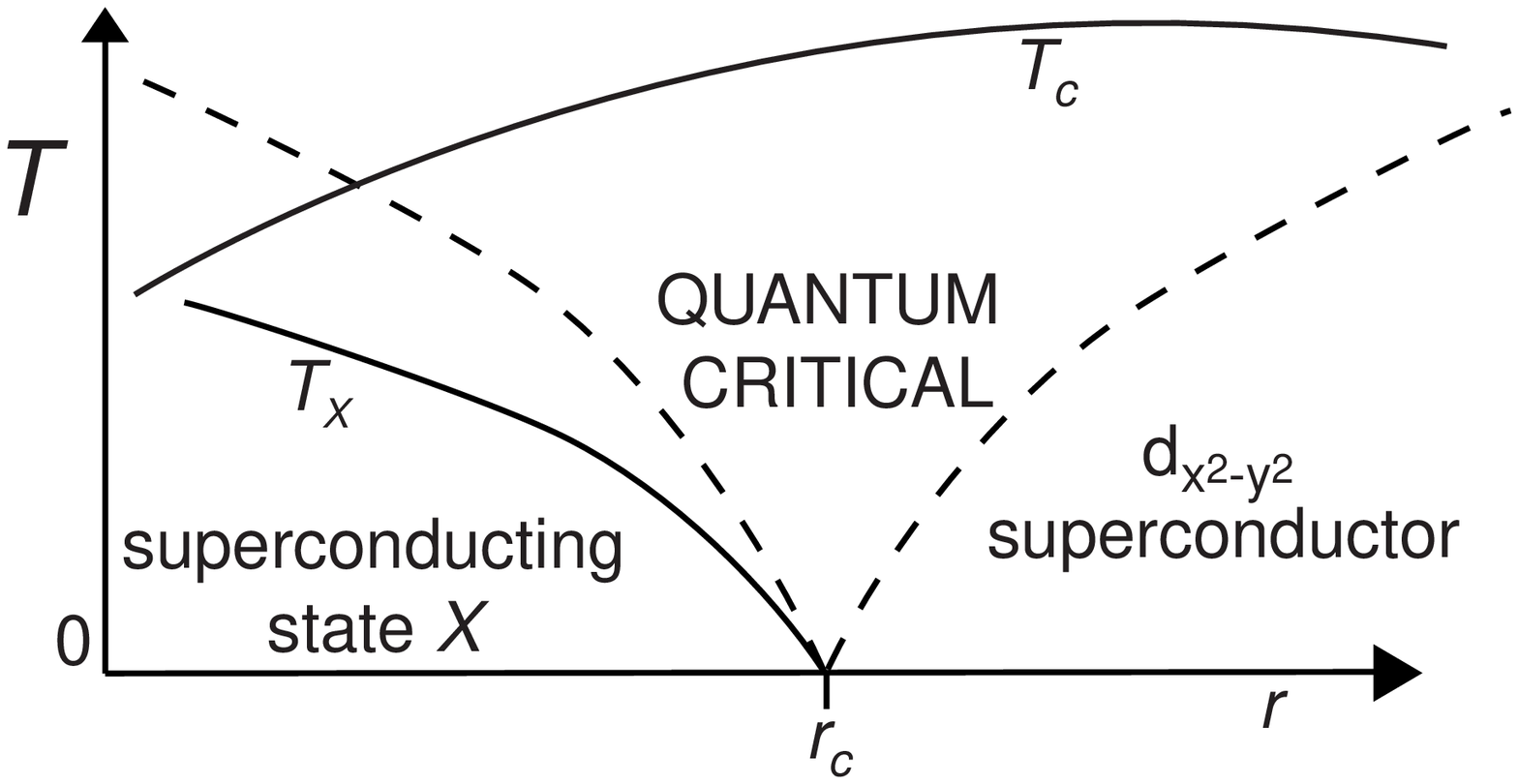} \caption{ Phase
diagram adapted from Ref~\protect\onlinecite{vzs}.
Superconductivity is present for $T<T_c$. The long-range order
associated with the state $X$ vanishes for $T>T_X$, but
fluctuations of this order provide anomalous damping of the nodal
quasiparticles in the quantum-critical region. } \label{fig1}
\end{figure}

Recent photoemission \cite{valla} and Thz conductivity\cite{joeo}
measurements on ${\rm Bi}_2 {\rm Sr}_2 {\rm Ca Cu}_2 {\rm
O}_{8+\delta}$, the cuprate superconductor, have indicated
anomalously large inelastic scattering of fermionic quasiparticles
near the gap nodes in the $d$-wave superconductor. While many
scattering mechanisms and scenarios have been proposed
\cite{gls,cmv,oned,sudip} for the damping of quasiparticles along
the $(1,0)$, $(0,1)$ directions (the ``anti-nodal
quasiparticles'') above the superconducting critical temperature
$T_c$, the possibilities below $T_c$ at the nodal points are much
more restricted, and allow us to make sharp distinctions between
competing theories. Standard BCS theory predicts a nodal
scattering rate $\sim T^3$ from short-range interactions, and
this is far too small to account for the observations. In this
paper we study a possible explanation\cite{vzs} due to
proximity to a quantum phase transition to some other
superconducting state $X$ (see Fig~\ref{fig1}). We show how
global symmetry and field-theoretic considerations permit a
classification of all possibilities for $X$, and we list those that
may account for the experiments.

The nodal quasiparticles at the gap nodes have a momentum
distribution curve (MDC) with a width proportional to $k_B T$
\cite{valla}, and there is little change\cite{jcc} in this
behavior when tuning $T$ through $T_c$.
The anti-nodal quasiparticles are broad and
ill-defined above $T_c$, but narrow dramatically below $T_c$,
forming long-lived states with an energy gap of 30-40 meV. A
natural possibility, based on other experimental probes
\cite{science}, is that these effects are due to proximity to a
quantum critical point to magnetic ordering. However, wavevector
matching conditions appear to rule this out for the nodal
quasiparticles: the magnetic fluctuations are strongest near
wavevector ${\bf Q}=(\pi,\pi)$, and while they can strongly
scatter anti-nodal quasiparticles above $T_c$, they do not
connect low energy quasiparticles near the nodes\cite{jcc}.

Rather than exploring the intricate details of the many
experiments, this paper performs the following well-posed
theoretical task: classify and describe theories in which a
$d$-wave superconductor at\cite{above} $T \ll T_c$ has, with
minimal fine-tuning, ({\em a}) a nodal quasiparticle MDC with a
width $\propto k_B T$, and possibly ({\em b}) negligible
scattering of the quasiparticles along $(1,0)$, $(0,1)$. We
find that theories which satisfy ({\em a}) also have a high
frequency tail \cite{addrem}
in the energy distribution curve (EDC) of the nodal
quasiparticles, as is experimentally observed \cite{valla,oned}.

Strong scattering of the gapless nodal quasiparticles surely
requires their coupling to some low-energy bosonic mode. It is
convenient to imagine that we have at our disposal some parameter
$r$ (which is possibly the hole concentration $\delta$, but not
necessarily so) which we can tune to condense the bosonic mode,
leading to a new superconducting state $X$ for $r < r_c$
(Fig~\ref{fig1}). The quantum-critical region of the phase
transition at $r=r_c$ and $T=0$ will satisfy ({\em a}) provided
the phase transition is below its upper critical dimension, and
the nodal fermions are intrinsic (in a sense to be made precise
below) degrees of freedom of the critical field theory \cite{SY}.
Conversely, ({\em b}) requires that the anti-nodal fermions are
merely spectators of the phase transitions, and are essentially
decoupled from the critical degrees of freedom.

\begin{figure}
\centerline{
\begin{tabular}{c|ccccc|c}
 & $E$ & $C_4^2$ & $ 2 C_4 $ & $ 2 I $ & $2 I^{\prime}$ & basis functions \\
\hline
$s$ & 1 & 1 & 1 & 1 & 1 & 1 \\
$p$ & 2 & -2 & 0 & 0 & 0 & $(\sin q_x , \sin q_y)$ \\
$d_{x^2 - y^2}$ & 1 & 1 & -1 & 1 & -1 & $\cos q_x - \cos q_y$ \\
$ d_{xy}$ & 1 & 1 & -1 & -1 & 1 & $\sin q_x \sin q_y$ \\
$g$ & 1 & 1 & 1 & -1 & -1 & $ \sin q_x \sin q_y ( \cos q_x - \cos q_y )$
\end{tabular}
}
\vspace{0.1in}
\caption{
Character table of the irreducible representations of the group $C_{4v}$. The $C_4$ rotations
are about the $z$ axis, and the $I$ ($I^{\prime}$) are reflections
about the $(1,0)$, $(0,1)$ ($(1,1)$, $(1,-1)$) directions; the
basis functions are chosen to be invariant under
translations by reciprocal lattice vectors.
}
\label{fig2}
\end{figure}

The most efficient scattering of nodal quasiparticles is provided
by a linear, non-derivative coupling between the fermion bilinears
and the order parameter; higher order and derivative couplings
have been considered recently \cite{vzs,nayak}, and invariably
lead \cite{vzs} to quasiparticle scattering rates that vanish with
super-linear powers of $T$. Order parameters which carry a net
momentum ${\bf Q}$, will, by momentum conservation, couple
linearly with the nodal fermions only if the spacing between two
of the nodal points is exactly ${\bf Q}$. Transitions involving
the onset of spin\cite{bfn} or site/bond charge density waves
\cite{vzs} (stripes) do satisfy \cite{vzs} ({\em a,b}) for
such values of ${\bf Q}$; however the restriction on ${\bf Q}$
could be a fine-tuning condition, and is not satisfied by the
${\bf Q}$ values observed so far. ``Staggered-flux''
order\cite{sudip,nayak} has a derivative coupling to the nodal
fermions, and ${\bf Q} = (\pi,\pi)$ which does not connect nodal
points: so ({\em a}) is not satisfied. Indeed, only the value
${\bf Q}=0$ naturally satisfies the constraints of momentum
conservation, and so we limit our attention to order parameters at
zero momentum. Furthermore, spin-triplet ordering at ${\bf Q}=0$
implies ferromagnetic correlations which are unlikely to be
present, and therefore we further restrict to spin-singlet
fermion bilinears. This means that our order parameter is a
component of the complex superconducting pairing function
$\Delta_{\bf q} = \langle c_{{\bf q} \uparrow} c_{-{\bf q}
\downarrow} \rangle$, or the  real excitonic (or `particle-hole')
pairing function $A_{{\bf q}} = \langle c_{{\bf q} a}^{\dagger}
c_{{\bf q} a} \rangle$ ($c_{{\bf q} a}$ annihilates an electron
with momentum ${\bf q}$ and spin $a=\uparrow , \downarrow$). It is
useful to decompose the functions $\Delta_{{\bf q}}$ and $A_{{\bf
q}}$ into components which transform under one of the irreducible
representations of the symmetry group of the Hamiltonian \cite{leggett}: this is
$C_{4v} \times Z_2$, where $C_{4v}$ is the tetragonal point group (see
Fig~\ref{fig2}), and the $Z_2$ component represents time-reversal
symmetry ${\cal T}$ (point group symmetry breaking
has been considered recently\cite{kfe,bss}, as have exciton
condensations \cite{nayak} at non-zero ${\bf Q}$).
Generically, a second-order transition can only occur by
condensation of an irreducible component
(multiple components can appear in successive transitions), and
this permits a complete classification of inequivalent order
parameters. Note that $d_{x^2-y^2}$ pairing is already present for
$r>r_c$ (see Fig~\ref{fig1}), and we will assume that
this ordering remains well-formed across the
transition; all our subsequent characterizations of possible orderings in
state $X$ refer to additional ordering beyond an implicitly
assumed background of $d_{x^2-y^2}$ pairing.
$A_{{\bf q}}$ is necessarily
even under ${\cal T}$, and so can generate $s,p,\ldots$ exciton
ordering; similarly $\Delta_{{\bf q}}$ can generate
$s,p,\ldots$ pairing or $is, ip \ldots$ pairing (the latter also
break ${\cal T}$), leading to a total of 15 possible order
parameters for $X$. Of these, $s$ exciton ordering is equivalent to an
innocuous
shift in the chemical potential, while $p$ and $ip$ pairing are
forbidden by Fermi statistics.
Because of the background $d_{x^2-y^2}$ pairing,
further $d_{x^2-y^2}$ or $id_{x^2-y^2}$
pairing is not a new ordering, while simple symmetry
considerations ({\em e.g.} examination of the fermion dispersion
relation in state $X$) show that $g$ excitons, $g$ pairing, and $d_{x^2-y^2}$
excitons are equivalent to $d_{xy}$ pairing, $d_{xy}$ excitons,
and $s$ pairing respectively. Only 7 inequivalent order parameters
now remain and we will discuss their properties shortly.

We begin by reviewing the action for low energy fermionic
excitations in a $d$-wave superconductor.
We denote the
components of $c_{{\bf q}a}$ in the vicinity of the four nodal
points $(\pm K, \pm K)$ ($K\approx 0.39 \pi$ at optimal doping)
by (anti-clockwise) $f_{1a}$, $f_{2a}$, $f_{3a}$, $f_{4a}$,
and introduce the 4-component Nambu spinors $\Psi_1 =
(f_{1a}, \varepsilon_{ab} f_{3b}^{\dagger})$
and  $\Psi_2 =
(f_{2a}, \varepsilon_{ab} f_{4b}^{\dagger})$ where
$\varepsilon_{ab}=-\varepsilon_{ba}$ and
$\varepsilon_{\uparrow \downarrow} = 1$.
Expanding to linear order in gradients from the nodal points,
we obtain
\begin{eqnarray}
S_{\Psi} &=& \int \!\!\! \frac{d^2 k}{(2 \pi)^2} T \!\sum_{\omega_n}
\Psi_1^{\dagger}  \left(
- i \omega_n + v_F k_x \tau^z + v_{\Delta} k_y \tau^x \right) \Psi_1  \nonumber \\
&~& \!\!\!\!\!\!\!\!\!\!\!\!\!\!\!\!\!\! +\int \!\!\! \frac{d^2 k}{(2 \pi)^2}
T \! \sum_{\omega_n}
\Psi_2^{\dagger}  \left(
- i \omega_n + v_F k_y \tau^z + v_{\Delta} k_x \tau^x \right) \Psi_2 .
\label{dsid1}
\end{eqnarray}
Here $\omega_n$ is a Matsubara frequency,
$\tau^{\alpha}$ are Pauli matrices which act in the fermionic
particle-hole space, $k_{x,y}$ measure the wavevector from the nodal points and
have been rotated
by 45 degrees from $q_{x,y}$ co-ordinates in Fig~\ref{fig2},
and $v_{F}$, $v_{\Delta}$
are velocities.

We now describe the 7 possible order parameters for state $X$,
along with the respective actions for the quantum phase
transition.
\newline
(A) \underline{$is$ pairing:} This has been considered in
Ref~\onlinecite{vzs}. The state $X$ (with $d_{x^2-y^2} + i s$
pairing) has no gapless fermionic excitations, breaks ${\cal T}$,
but all charge neutral observables (like the charge density or
lattice displacements) retain the full $C_{4v}$ symmetry. The order
parameter transforms as a real, one-dimensional representation of
$C_{4v} \times Z_2$, and so can be represented by a single, real
field $\phi$; this will also be true for (B)-(F) below, with
only (G) requiring a doublet of real fields. On general
symmetry grounds, following action for $\phi$ is obtained after
integrating out high energy fermion modes:
\begin{equation}
S_{\phi} = \int \!\! d^2 x d \tau \Big[
\frac{1}{2}(\partial_{\tau} \phi)^2 + \frac{c^2}{2} (\nabla \phi )^2 +
\frac{r}{2} \phi^2 + \frac{u}{24} \phi^4 \Big];
\label{dsid3}
\end{equation}
here $\tau$ is imaginary time,
$c$ is a velocity, $r$ tunes the system across the
quantum critical point, and $u$ is a quartic self-interaction. By
itself, $S_{\phi}$ would describe a critical point at $r=r_c$ in
the universality class of the classical, three-dimensional Ising
model. However, a coupling to the low energy
fermionic modes in (\ref{dsid1}) can preempt this conclusion \cite{bfn}:
its form can be deduced from the values of the basis function in
Fig~\ref{fig2} at the nodal points, and the information that the
order parameter is in the particle-particle channel--
\begin{equation}
S_{\Psi\phi} = \int \!\! d^2 x d \tau \Big[ \lambda \phi
\left( \Psi_1^{\dagger} M_1 \Psi_1 + \Psi_2^{\dagger} M_2
\Psi_2 \right) \Big],
\label{dsid4}
\end{equation}
where $\lambda$ is the required linear coupling constant between the order
parameter and a fermion bilinear, and $M_1=M_2=\tau^y$.
\newline
(B) \underline{$i d_{xy}$ pairing}: This is very similar to $A$, with the main
change arising from the new basis function in Fig~\ref{fig2},
which now implies $M_1=-M_2 = \tau^y$.
\newline
(C) \underline{$i g$ pairing}: Also related to (A), but now the basis function
in Fig~\ref{fig2} vanishes at the nodal points.
Consequently, the coupling between $\Psi$ and $\phi$ requires at
least one spatial derivative, and is irrelevant \cite{vzs}.
The action $S_{\phi}$ in (\ref{dsid3}) is the entire critical
theory of the transition, and the scattering of the nodal fermions
is weak, arising only from irrelevant couplings, and violates
 ({\em a}).
\newline
(D) \underline{$s$ pairing}: ${\cal T}$ remains unbroken, but the
symmetry of charge neutral observables
is broken to $C_{2v}$, so that $X$ (with $d_{x^2-y^2} + s$ pairing)
is a superconducting
nematic \cite{kfe,bss}. The nematic order is polarized along the $(1,0)$
or $(0,1)$ directions. For weak ordering,
the state $X$ retains gapless nodal fermionic excitations, but
the nodal points are at $(\pm K^{\prime}, \pm K)$ with $K^{\prime} \neq K$;
for a sufficiently large $s$ component, the nodal points
disappear
upon colliding in pairs as $\mbox{min} (K^{\prime}, K) \rightarrow 0$,
in a separate quantum critical point which is not of
interest here.
As in (A,B), coupling of the order
parameter is described by (\ref{dsid4}), but with
$M_1=M_2=\tau^x$.
\newline
(E) \underline{$d_{xy}$ excitons}: This is as in (D), but symmetry of charge neutral observables
in $X$ is broken
 to a different $C_{2v}$ subgroup of $C_{4v}$, with the nematic
now polarized along the diagonal $(1, \pm 1)$ directions.
The nodal points in $X$ are at $\pm (K,K)$ and $\pm (K^{\prime}, -K^{\prime})$
 with $K \neq K^{\prime}$.
In the action (\ref{dsid4}), we now have
$M_1=-M_2=\tau^z$.
\newline
(F) \underline{$d_{xy}$ pairing}: Such an ordering in $X$ moves the nodal
points clockwise (or anti-clockwise) from $(\pm K, \pm K)$,
reducing the $C_{4v}$ symmetry
to $C_4$, while preserving ${\cal T}$.
Again the action (\ref{dsid4})
describes the order parameter/fermion coupling, but with $M_1 = -M_2 = \tau^x$.
\newline
(G) \underline{$p$ excitons}: The order parameter transforms under a
two-dimensional representation of $C_{4v}$, requiring a doublet of
real fields, $(\phi_x, \phi_y)$, to describe the low energy
bosonic modes. The state $X$ retains ${\cal T}$ and the gapless
nodal points, but has $C_{4v}$ broken to $Z_2$. The action (\ref{dsid3}) is
replaced by
\begin{eqnarray}
\widetilde{S}_{\phi} &=& \int \!\! d^2 x d \tau \left[\frac{1}{2}
\left\{
(\partial_{\tau} \phi_x)^2 + (\partial_{\tau} \phi_y)^2 +
c_1^2 (\partial_x \phi_x )^2 \right. \right. \nonumber \\
+ &c_2^2 & (\partial_y \phi_x )^2 +
c_2^2 (\partial_x \phi_y )^2 + c_1^2 (\partial_y \phi_y )^2 +
e (\partial_x \phi_x ) (\partial_y \phi_y) \nonumber \\
&+& \left. \left. r(\phi_x^2 + \phi_y^2) \right\}  + \frac{1}{24}
\left\{u (\phi_x^4 + \phi_y^4) + 2 v \phi_x^2 \phi_y^2 \right\}
\right],
\label{p1}
\end{eqnarray}
while the coupling between $\phi_{x,y}$ and $\Psi_{1,2}$ is
\begin{equation}
\widetilde{S}_{\Psi\phi} = \int \!\! d^2 x d \tau \Big[ \lambda
\left(\phi_x \Psi_1^{\dagger} \Psi_1 + \phi_y \Psi_2^{\dagger}
\Psi_2 \right) \Big].
\label{p2}
\end{equation}

We now make a few general remarks on the field theories above.
Upon integrating out the fermion fields, we find a finite
one-loop renormalization of the tuning parameter, $r$. This should
be contrasted with the behavior in a system with a Fermi surface,
where we would find the BCS infrared logarithmic divergence
in the analogous term: this is, of
course,
the reason that a $T=0$
Fermi liquid is unstable to superconductivity for any
attractive interaction. In
the present situation, the background $d_{x^2-y^2}$
superconductivity has reduced the Fermi surface to 4 Fermi points,
and so further pairing or excitonic
instabilities occur at finite values of $r$ and
$\lambda$. Indeed, this feature allows
a non-trivial quantum critical point, with a universal
quantum-critical region (Fig~\ref{fig1}); the
fluctuations in this region will satisfy ({\em a}) provided
the quantum-critical point at $r=r_c$, $T=0$ is described by a
fixed point of the renormalization group (RG) transformation at which $\lambda$
approaches a non-zero and finite fixed point value---then the
scattering rate of the nodal fermions will be determined by $T$
alone \cite{book}.

The results of our RG analysis of (A--G) are simple and
remarkable. Only for (A,B,C) do we find a fixed point, accessed by
tuning the parameter $r$; such a fixed point describes a
second-order quantum phase transition at the critical point
$r=r_c$. For all other cases, we find runaway flows of the
couplings, with no non-trivial fixed points, which suggests
first-order transitions. As we have already noted, the fixed
point for (C) is the Ising model---the nodal fermions are
decoupled from the critical degrees of freedom in the scaling
limit, so that ({\em a}) is not satisfied. Only (A) and (B)
satisfy ({\em a}), with the couplings $\lambda$ and $u$
approaching non-zero fixed point values: the nodal fermions and
$\phi$ are strongly coupled in the critical theory, and the
anomalous dimension of the fermion field leads to a large
$\omega$ tail in its EDC \cite{vzs,book}. The (A, B) fixed points
are also Lorentz invariant---the dynamic exponent $z=1$, and the
velocities renormalize to $v_F = v_{\Delta} =c$ in the scaling
limit. Indeed, these fixed points were discussed earlier
\cite{vzs}, but only for almost equal velocities; here we have
established that the equal-velocity fixed point is the only one
for arbitrary initial velocities. However, the crossover exponent
which determines how rapidly the velocities approach each other
is negligible \cite{vzs} ($\approx 0.05$), so that a transient
regime with unequal velocities will be realized over essentially
all of the experimentally accessible regime.

The methodology of our RG is standard and details appear
elsewhere \cite{rome}. The familiar momentum-shell method, in
which degrees of freedom with momenta between $\Lambda$ and
$\Lambda-d\Lambda$ are successively integrated out, fails here:
the combination of momentum dependent renormalizations at one
loop, the direction-dependent velocities ($v_F$, $v_{\Delta}$,
$c$ \ldots), and the hard momentum cutoff generate unphysical
non-analytic terms in the effective action. So we obtained the RG
equations by 
using a soft cutoff at scale $\Lambda$,
and by taking a $\Lambda$ derivative of the renormalized vertices
and self energies. We obtained equations for all the velocities,
the dynamic exponent $z$, and the field anomalous dimensions to
one-loop order in the non-linearities $\lambda$, $u$, $v$. For
(D,E,F) a simple and robust effect preempts a fixed point: the
structure of $M_{1,2}$ produces opposite sign renormalizations
for $v_{F,\Delta}$, in a manner that both flow equations cannot
simultaneously be at a fixed point; (G) required a more detailed
analysis.

Our main result is that, among the 7 transitions considered here,
only for those involving onset of $d_{x^2-y^2}+is$ or
$d_{x^2-y^2}+id_{xy}$ pairing in a $d_{x^2-y^2}$ superconductor
did we find a universal critical theory of coupled fermionic and
bosonic order parameter modes below its upper critical dimension.
Such transitions naturally satisfy ({\em a}). Upon further
imposing condition ({\em b}), case (B), with $d_{x^2-y^2} + i
d_{xy}$ pairing, is uniquely selected: from the basis functions
in Fig~\ref{fig1} we see that $\phi$ couples to fermions in all
directions for (A), while the fermionic coupling vanishes along
the anti-nodal directions for (B)---so the gapped anti-nodal
fermions will [will not] lose the sharp quasi-particle peak below
$T_c$ by emission of multiple $\phi$ quanta, for (A) [(B)].

Pairing in the $d_{x^2-y^2}+i d_{xy}$ channel has been
considered in numerous works recently
\cite{sigrist}, with the order in the
ground state either global (induced spontaneously or by an
external magnetic field) or local (in the vicinity of
defects\cite{kirtley}, surfaces\cite{surfaces},
or vortices\cite{dhlee}). Here we only require strong fluctuations of such order,
induced by a proximity to a hypothetical point in the phase diagram
where global order arises. While experimental discovery of such a point
is of course preferable, tests of our
proposal would also be provided by signals of $\phi$ fluctuations.
This is a spin-singlet mode with $d_{xy}$
symmetry, odd under time-reversal, and at $T=0$ it
has spectral weight
with mean frequency and width both of order of an
energy scale $\sim (r-r_c)^{z\nu}$ (where $\nu$ is the usual
correlation length exponent)---we estimate this scale is $\sim 5-10$ K;
in the quantum-critical region the characteristic energy scale
is $k_B T/\hbar$. Fluctuations of $\phi$ should lead to anomalies
in Raman scattering \cite{ss} and Hall transport \cite{ong}:
these issues will be discussed in future work.

We thank P.~Aebi, D.~Baeriswyl, L.~Balents, J.~Campuzano,
L.~Cooper, S.~Kivelson, A.~Millis, C.~Morais-Smith, S.~Shastry and
J.~Zaanen for useful discussions, and
D.~Baeriswyl for
hospitality at the Universit\'{e} de Fribourg.
This research was
supported by US NSF Grant No DMR 96--23181 and by the DFG (VO
794/1-1).

\newpage
\begin{widetext}
\begin{center}
~\\
~\\
{\large\bf Erratum: Quantum phase transitions in $d$-wave superconductors\\
Phys. Rev. Lett. {\bf 85}, 4940  (2000)}\\
~\\
Matthias Vojta\\
{\it Institut f\"ur Theoretische Physik,
Universit\"at zu K\"oln, Z\"ulpicher Str. 77, 50937 K\"oln, Germany}\\
~\\
{Ying Zhang}\\
{\it Goldman Sachs, 1 New York Plaza, New York, NY 10004}\\
~\\
{Subir Sachdev}\\
{\it Department of Physics, Harvard University, Cambridge,
MA 02138}\\
~\\
We correct an error in our paper Phys. Rev. Lett. {\bf 85}, 4940 (2000)
[arXiv:cond-mat/0007170].\\
Our characterization of the physical properties of the
superconducting state G was incorrect:
it\\ breaks time-reversal symmetry, carries spontaneous currents,
and possesses Fermi surface pockets.
\end{center}
~\\
~\\

\end{widetext}

In the discussion of case G in Ref.~\onlinecite{prl}, above Eq.~(4), the single 
sentence
``The state $X$ retains $\mathcal{T}$ and the gapless nodal
points, but has $C_{4v}$ broken to $Z_2$'' is incorrect.
The state $X={\rm G}$ breaks $\mathcal{T}$ (time-reversal), and has spontaneous
electrical currents. For $\phi_x \neq 0$ and $\phi_y = 0$ (or vice versa) the 
currents
have the same symmetry as those in the state $\Theta_{\rm II}$
discussed by Simon and Varma \cite{varma2}.
Also, as pointed out by Berg {\em et al.} \cite{berg}, the nodal
quasiparticles do not survive in the superconducting state G, but turn into 
Fermi pockets.
The latter conclusion can be verified from the fermion spectrum
obtained by diagonalizing Eqs. (1)+(5) for constant $\phi_{x,y}$.

All other sentences and the conclusions in the paper \cite{prl} remain 
unchanged.

Also, in the companion paper, Ref.~\onlinecite{long}, the only error is in
the sketch of the
fermion excitations in Fig.~2 for case G.

We thank Erez Berg and Cenke Xu for helpful discussions.
This research was supported by NSF grant DMR-0537077
and DFG SFB 608.


\begin{references}


\bibitem{valla} T.~Valla {\em et al.}, Science {\bf
285}, 2110 (1999).

\bibitem{joeo} J.~Corson, J.~Orenstein, and J.~N.~Eckstein,
Phys. Rev. Lett. {\bf 85}, 2569 (2000).

\bibitem{gls} M.~Franz and A.~J.~Millis, Phys. Rev. B {\bf 58}, 14572 (1998);
H.-J.~Kwon and A.~T.~Dorsey, Phys. Rev. B {\bf 59}, 6438 (1999);
V.~P.~Gusynin, V.~M.~Loktev, and S.~G.~Sharapov, Sov. Phys. JETP
{\bf 90}, 993 (2000); P.~Monthoux and D.~Pines, Phys. Rev. B {\bf
49}, 4261 (1994) and {\bf 50}, 16015 (1994); A.~Paramekanti {\em
et al.}, Phys. Rev. B {\bf 62}, 6786 (2000).

\bibitem{cmv} C.~M.~Varma,
Phys. Rev. Lett. {\bf 83}, 3538 (1999);
L.~B.~Ioffe and A.~J.~Millis, Phys. Rev. B {\bf 58},
11631 (1998); S.~Caprara {\em et al.}, Phys. Rev. B {\bf 59},
14980 (1999); T.~Senthil and M.~P.~A.~Fisher,
\prb {\bf 62}, 7850 (2000) ; cond-mat/9912380.

\bibitem{oned} D.~Orgad {\em et al.}, cond-mat/0005457.

\bibitem{sudip} S.~Chakravarty {\em et al.}, cond-mat/0005443.

\bibitem{vzs} M.~Vojta, Y.~Zhang, and S.~Sachdev, Phys. Rev. B
{\bf 62}, 6721 (2000).

\bibitem{jcc} A.~Kaminski {\em et al.}, Phys. Rev. Lett. {\bf 84}, 1788
(2000); cond-mat/0004482. Below $T_c$, there is a change in the Fermi velocity at
an energy $\sim 80$ meV; this feature appears to be well-explained
by a coupling to a collective spin resonance mode
[Ar.~Abanov and A.~V.~Chubukov, Phys. Rev. Lett.
{\bf 83}, 1652 (2000); M.~Eschrig and M.~R.~Norman,
\prl {\bf 85}, 3261 (2000)]
and will not be directly
relevant to the theory presented here.

\bibitem{science} S.~Sachdev, Science {\bf 288}, 475 (2000).

\bibitem{above} The same theories should also apply above $T_c$
at length scales shorter than the superconducting
phase coherence length (S.~Sachdev and M.~Vojta, cond-mat/0005250).

\bibitem{addrem} For frequencies $\hbar \omega \gg k_B T$, the region of nonzero
spectral weight is defined by $\omega > c k$ \cite{vzs}, where $k$ is the wave vector
and $c$ is a velocity, as is also the case in one-dimensional models
\cite{oned}.

\bibitem{SY} S.~Sachdev and J.~Ye, Phys. Rev. Lett. {\bf 69}, 2411 (1992).

\bibitem{nayak} C.~Nayak, Phys. Rev. B {\bf 62}, 4880 (2000).

\bibitem{bfn} L.~Balents, M.~P.~A.~Fisher, and C.~Nayak, Int. J. Mod.
Phys. B {\bf 12}, 1033 (1998).

\bibitem{leggett} J.~Annett, N.~Goldenfeld, and A.~J.~Leggett,
{\em Physical Properties of High Temperature Superconductors,} Vol. 5,
D.~M.~Ginsberg (ed.), World Scientific, Singapore (1996).

\bibitem{kfe} S.~A.~Kivelson, E.~Fradkin, and V.~J.~Emery, Nature
{\bf 393}, 550 (1998). They
formulate their theory for the nematic by departing from an extreme one-dimensional limit,
whereas we depart from the two-dimensional $d$-wave superconductor.

\bibitem{bss} B.~S.~Shastry and T.~V.~Ramakrishnan, unpublished.

\bibitem{book} S.~Sachdev, {\em Quantum Phase Transitions},
Cambridge University Press, Cambridge (1999).

\bibitem{rome} M.~Vojta, Y.~Zhang, and S.~Sachdev,
cond-mat/0008048.

\bibitem{sigrist} See the review M.~Sigrist, Prog. Theor. Phys.
{\bf 99}, 899 (1998).

\bibitem{kirtley} F.~Tafuri and J.~R.~Kirtley,
cond-mat/0003106.

\bibitem{surfaces} M. Covington {\em et al.},
Phys. Rev. Lett. {\bf 79}, 277 (1997);
S.~Kos cond-mat/0006047.

\bibitem{dhlee}  M.-R.~Li, P.~J.~Hirschfeld, and P.~Woelfle,
cond-mat/0003160; J.~H.~Han and D.-H.~Lee, Phys. Rev. Lett. {\bf
85}, 1100 (2000).

\bibitem{ss} B.~S.~Shastry and B.~I.~Shraiman, Phys. Rev. Lett.
{\bf 65}, 1068 (1990); S.~Yoon {\em et al.}, \prl {\bf 85}, 3297 (2000).

\bibitem{ong} K.~Krishana {\em et al.}, Phys. Rev. Lett. {\bf 82},
5108 (1999).


\end{references}

\begin{thebibliography}{99}

\bibitem{prl}
M. Vojta, Y. Zhang, and S. Sachdev,
Phys. Rev. Lett. {\bf 85}, 4940  (2000).

\bibitem{varma2} M. E. Simon and C. M. Varma, \prl {\bf 89}, 247003 (2002),
see Fig 1b.

\bibitem{berg} E.~Berg, C.-C.~Chen, and S.~A.~Kivelson,
Phys. Rev. Lett. {\bf 100}, 027003 (2008).

\bibitem{long} M.~Vojta, Y.~Zhang, and S.~Sachdev,
Int. J. Mod. Phys. B {\bf 14}, 3719 (2000) [arXiv:cond-mat/0008048].

\end{thebibliography}
\end{document}